\documentclass[12pt]{elsarticle}

\usepackage{graphicx,epstopdf} 
\usepackage{amsmath,amssymb,latexsym} 
\usepackage{hyperref} 
\usepackage{parskip}
\usepackage{float}
\usepackage[font={small}]{caption}
\usepackage{subcaption}
\usepackage{mathtools}
\usepackage{indentfirst}
\usepackage{titlesec}

\titleformat{\section}
  {\normalfont\fontsize{14}{20}\sffamily\bfseries}
  {\thesection}
  {1em}
  {}

\titleformat{\subsection}
  {\normalfont\fontsize{12}{17}\sffamily\bfseries\slshape}
  {\thesubsection}
  {1em}
  {}
  
\title{Characterization of the Hamamatsu 8'' R5912-MOD Photomultiplier Tube}
\author{Tanner Kaptanoglu$^{a}$ \footnote{$^{*}$ Corresponding Author \\ \hspace{5mm} Email Address: tannerk@hep.upenn.edu} \\ \vspace{3mm}
\textit{$^{a}$University of Pennsylvania, Philadelpha PA 19104, USA}}

\begin{document}

\pagestyle{plain} 


\begin{abstract}
\par Current and future neutrino and direct detection dark matter experiments hope to take advantage of improving technologies in photon detection. Many of these detectors are large, monolithic optical detectors that use relatively low-cost, large-area, and efficient photomultiplier tubes (PMTs). A candidate PMT for future experiments is a newly developed prototype Hamamatsu PMT, the R5912-MOD. In this paper we describe measurements made of the single photoelectron time and charge response of the R5912-MOD, as well as detail some direct comparisons to similar PMTs. Additionally, a 1D scan of the photocathode and an after pulsing measurement were performed. Most of these measurements were performed on three R5912-MOD PMTs operating at gains close to $1 \times 10^{7}$. The transit time spread ($\sigma$) and the charge peak-to-valley were measured to be on average 680ps and 4.2 respectively. The results of this paper show the R5912-MOD is an excellent candidate for future experiments in several regards, particularly due to its narrow spread in timing.

\end{abstract}

\maketitle
\tableofcontents

\section{Introduction}

Large optical detectors used in neutrino and direct dark matter experiments often rely on PMTs for efficient detection of scintillation and Cherenkov light. The Super-Kamiokande experiment  \cite{superk} used a 40 kilo-tonne water Cherenkov detector to provide evidence of atmospheric neutrino oscillations. The Sudbury Neutrino Observatory (SNO) \cite{SNO} used a 1 kilo-tonne heavy water detector to measure $^{8}$B solar neutrinos and in turn definitively solve the solar neutrino problem. Scintillator detectors such as KamLAND \cite{KAMLAND}, RENO \cite{RENO}, and Daya Bay \cite{DAYABAY} have used measurements of reactor neutrinos to determine neutrino oscillation parameters. Borexino \cite{Borexino} uses a 280 tonne scintillator experiment purposed for low energy solar neutrinos. Many direct dark matter experiments, including LUX \cite{LUX} and DEAP \cite{DEAP}, detect the scintillation light emitted during a hypothetical dark matter interaction in their liquid Xenon or liquid Argon filled detectors. Every one of these detectors takes advantage of large-area PMTs.

The physics goals of these experiments is in part determined by the performance of the PMTs. Optimal PMTs include features such as large collection areas, high efficiencies, small timing jitters, and narrow charge resolutions. Better performing PMTs can enhance the performance of reconstruction algorithms, increase background rejection and improve energy resolution. Potential future experiments such as Hyper-Kamiokande \cite{hyperk}, NuPRISM \cite{nuprism}, and THEIA \cite{theia} might benefit from taking advantage of the R5912-MOD PMTs. 

The R5912-MOD is an ideal candidate particularly because of its relatively large detection area, its narrow spread in transit time, and its excellent charge resolution. Measurements described in Sections \ref{sec:charge} and \ref{sec:timing} show state-of-the-art single photoelectron response. In Section \ref{sec:eff} we directly compare this response to similar PMTs in the same experimental setup. A 1-dimensional scan across the photocathode is performed to compare the response across the PMT and is described in Section \ref{sec:1d}. An after pulsing measurement designed to probe the very late time PMT pulses coming from drifting ions in the PMT vacuum is descried in Section \ref{sec:afterpulsing}. Hamamtasu can make more of the R5912-MOD PMTs for sale \cite{private}. 

\section{R5912-MOD}\label{sec:overview}

\begin{figure}[ht]
\centering
\includegraphics[trim={0cm 0cm 0cm 0.5cm}, clip = true, scale=0.5]{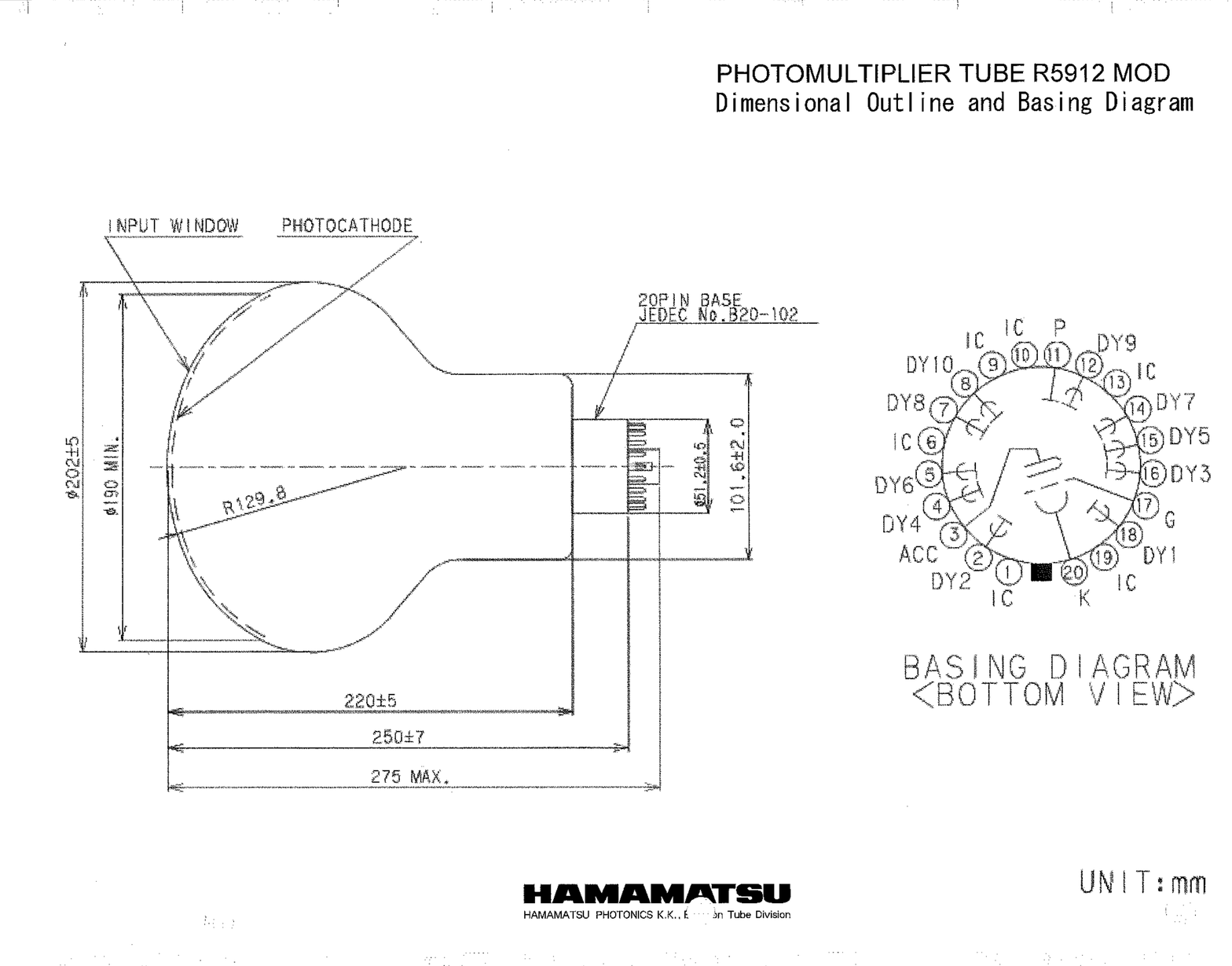}
\caption{The R5912-MOD PMT specifications, including the pin-out for the base diagram, as provided by Hamamtasu.}
\label{fig:pmt}
\end{figure}

\begin{figure}[ht]
\centering
\includegraphics[scale=0.3, angle=180]{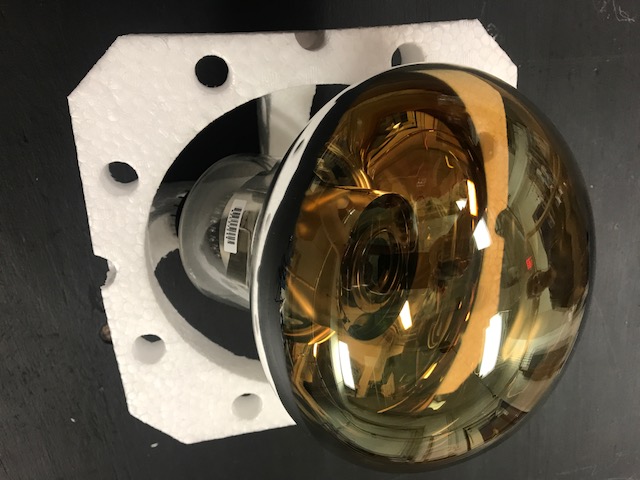}
\includegraphics[scale=0.3, angle=180]{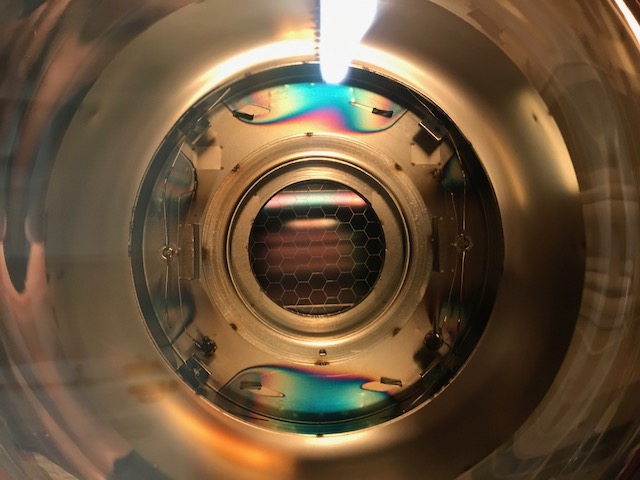}
\caption{An image of the full R5912-MOD PMT (left) and a photo of the inside of the PMT (right).}
\label{fig:pmt-pic}
\end{figure}

The R5912-MOD is a 8-inch PMT developed by Hamamtasu Photonics with 10 linearly focused dynode stages. Shown in Figure \ref{fig:pmt} is the PMT specifications showing the dimensions, photocathode area, and basing diagram. In Figure \ref{fig:pmt-pic} are pictures of the PMTs which visually show the optics of the photocathode. In this paper we tested three of the R5912-MOD PMTs and, in several cases, compared their characteristics to similar PMTs. These prototype PMTs are expected to be less efficient than Hamamatsu's R5912-100 PMTs, which peak around 35\% quantum efficiency (QE); however Hamamatsu could incorporate the super biakali (HQE) photocathode on these PMTs \cite{private}.

\subsection{PMT Base Testing}\label{sec:base}

The PMT base is the circuitry associated with providing the voltage to each dynode stage. In addition, the base also shapes and sizes the PMT pulse and thus impacts the PMTs response. We tested several different base designs for the R5912-100 PMT, each time attempting to optimize the performance of the PMT. In particular, the design chosen had the smallest spread in the prompt transit time of the PMT, the measurement for which is described in detail in Section \ref{sec:timing}. The voltage divider ratios for the chosen base is shown in Table \ref{tab:dy1}. The base design also determines the high voltage that needs to be supplied in order to achieve a gain of $1 \times 10^{7}$, and was tuned so that the PMT operated at around 1800V. The measurements described in this paper are all for this particular base design.

\begin{table}[ht]
\footnotesize
\centering
\begin{tabular}{|c|c|c|c|c|c|c|c|c|c|c|c|c|c|c|} \hline
- & K & DY1 & G & DY2 & DY3 & DY4 & DY5 & DY6 & DY7 & ACC & DY8 & DY9 & DY10 & P \\ \hline \hline
Res. (ratio) & 11.5 & 1 & 3.5 & 4 & 2 & 2 & 1 & 1 & 0 & 1 & 1 & 1 & 1 & \\ \hline
Cap ($\mu$F) & & & & & & & & & & & 0.01 & 0.01 & 0.01 & \\ \hline 
\end{tabular}
\caption{Voltage divider ratios between each stage. For example the ratio for the drop from the cathode to the first dynode is 11.5 times larger than the voltage drop between the first dynode and the grid. The total resistance across the entire base is approximately 16.8M$\Omega$.}
\label{tab:dy1}
\end{table}

\begin{figure}[ht]
\centering
\includegraphics[scale=0.3, angle=270]{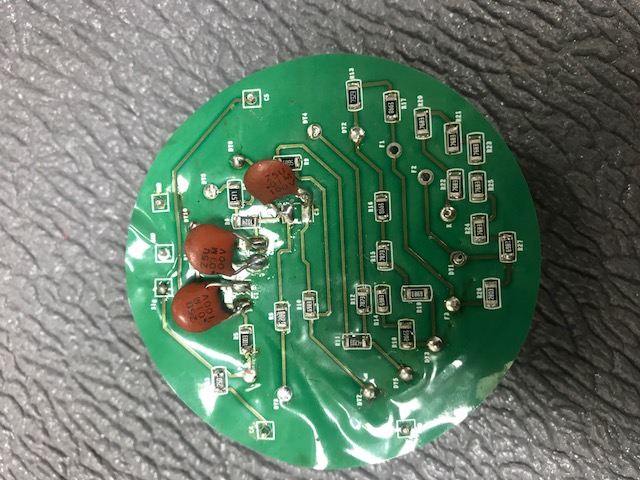} \hspace{10mm}
\includegraphics[scale=0.3, angle=270]{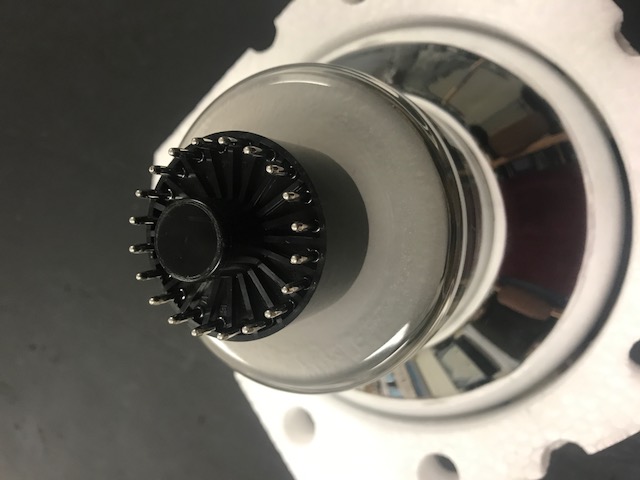}
\caption{A picture of the finalized PMT base (left). Not seen are the HV capacitors, which are on the other side of the base. The socketed PMT is shown on the right.}
\label{fig:baseandsocket}
\end{figure}

\section{Single Photoelectron Characterization (SPE)}\label{sec:SPE}

A PMT's single photoelectron response is critical to its performance. Many large water Cherenkov, liquid scintillator, and dark matter detectors expect the deployed PMTs to detect primarily SPEs. Any such detector is improved by increasing the number of detected photons, very accurately measuring the photon arrival times (hit times), and being able to accurately separate SPE from multi-PE (MPE) PMT hits. In order to achieve these goals these detectors often utilize large arrays of PMTs, hopefully with excellent charge and timing characteristics.

The SPE charge distribution is important for several reasons. The shape of the SPE charge distribution, in particular, the width of the primary SPE peak and the lack of a low or high charge tail, allows for higher detection efficiency and improved photon counting. These gains in efficiency are important for optical detectors where energy resolution is determined by the number of detected photoelectrons.

The SPE transit time distribution is a critical feature of a PMTs response. The PMT hit times are used in almost any reconstruction algorithm hoping to determine the position of an event in the detector. The more accurately one can measure the time of the PMT hit, the better these algorithms perform. Additionally, many future scintillator detectors hope to separate the Cherenkov light component from the scintillation component using the prompt nature of Cherenkov light. However, PMTs with broad transit time spreads make this difficult to accomplish \cite{timing}. As is shown in \cite{orebigann} the separation of Cherenkov and scinillation light has been demonstrated using 1'' PMTs, with very fast timing. As will be discussed in Section \ref{sec:timing} the R5912-MOD PMT has an extremely narrow spread in timing for a PMT of its size which might make it possible to separate the Cherenkov and scintillation components using a large area PMT. 

\subsection{Experimental Setup}\label{sec:setup}

\begin{figure}[ht]
\centering
\includegraphics[scale=0.4]{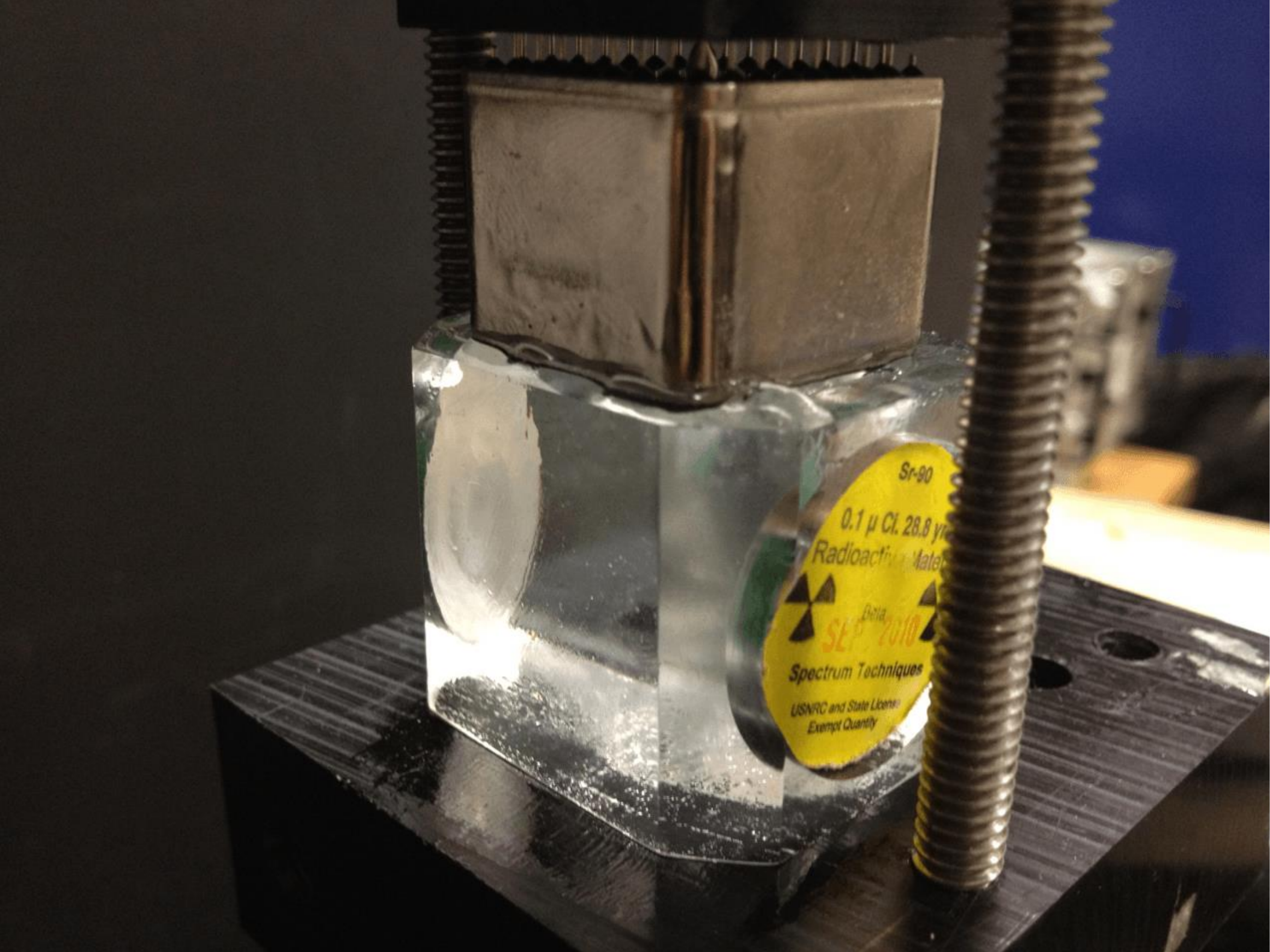}
\caption{The UV transparent acrylic embedded with two $^{90}$Sr sources is optically coupled to an R7600-200 1-inch cubic PMT. This apparatus is referred to as the Cherenkov source in this paper.}
\label{fig:source}
\end{figure}

The experimental setup to measure the SPE characterization takes advantage of a fast trigger PMT and a Cherenkov light source in order to provide a source of single photons to the R5912-MOD PMT. The Cherenkov source consists of an acrylic block embedded with two plastic $^{90}$Sr disk sources. The $^{90}$Sr undergoes a 0.546 MeV $\beta^{-}$ decay to $^{90}$Y with a half life of 29.1 years. The $^{90}$Y undergoes a 2.28MeV $\beta^{-}$ decay to $^{90}$Zr with a half-life of 64 hours. The $\beta^{-}$s from both decays enter the acrylic and create Cherenkov light. The emitted Cherenkov light has the advantage of being produced with an extremely narrow spread in timing. Additionally, the wavelength spectrum of Cherenkov light is well-known and spans the same spectrum as many common scintillators. Finally, the Cherenkov process produces relatively few photons per interaction, making it easy to move the R5912-100 far enough away from the source to primarily see single photons. The acrylic in the Cherenkov source is UV-transparent and is the same acrylic used in the construction of the SNO acrylic vessel. The acrylic block is optically coupled with Saint-Gobain BC-630 optical grease to a fast trigger PMT, a 1-inch cubic R7600-200 super bialkali high quantum efficiency (HQE) Hamamatsu PMT, which has a transit time spread of roughly 250ps FWHM and a quantum efficiency peaking around 40\%. The very narrow spread in transit time and the lack of late-pulsing and pre-pulsing is critical for the trigger PMT. The acrylic source embedded with the plastic $^{90}$Sr sources as well as the trigger PMT with the optical couping gel are shown in Figure \ref{fig:source}.

\begin{figure}[ht]
\centering
\includegraphics[scale=0.8, trim=1cm 1cm 0cm 4cm, clip=true]{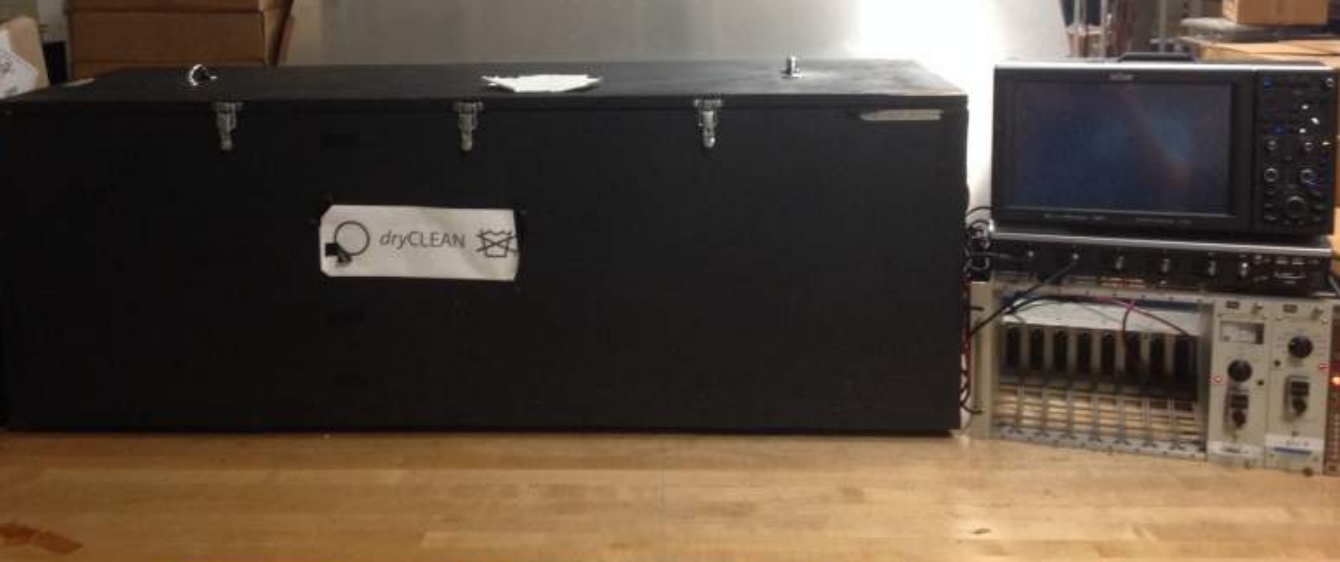}
\caption{The darkbox, high voltage supply, and oscilloscope used for data taking.}
\label{fig:darkbox}
\end{figure}

The R5912-MOD PMT is kept 30cm away from the Cherenkov source in order to maintain a primarily SPE source of light. Additionally, this distance ensures the entire front-face of the PMT is illuminated. This setup is housed in a darkbox shown in Figure \ref{fig:darkbox} and the high voltage is provided to each PMT by the ISEG NHS supply, a high precision, six-channel NIM module. The ripple and noise on the supply are much less than 1V, making the supply ideal for powering PMTs. The darkbox is lined with Finemet magnetic shielding in order to minimize interference from the Earth's magnetic field.

A Lecroy WaveRunner 606Zi 600MHz oscilloscope is used to digitize the signals from both the R7600-200 and the R5912-MOD PMT. The signal from the R7600-200 PMT is used to trigger the oscilloscope readout, and the PMT is often simply referred to as the trigger PMT in this paper. For the SPE measurements the waveforms extracted were 500.2ns long using 50ps samples. The scope has an 8-bit ADC with a variable dynamic range, which allowed for roughly 300$\mu$V vertical resolution. The LeCrunch \cite{lecrunch} software was used to readout the data from the scope over ethernet as well as format the data into custom HDF5 files. 

\begin{figure}[ht]
\centering
\includegraphics[scale=0.8]{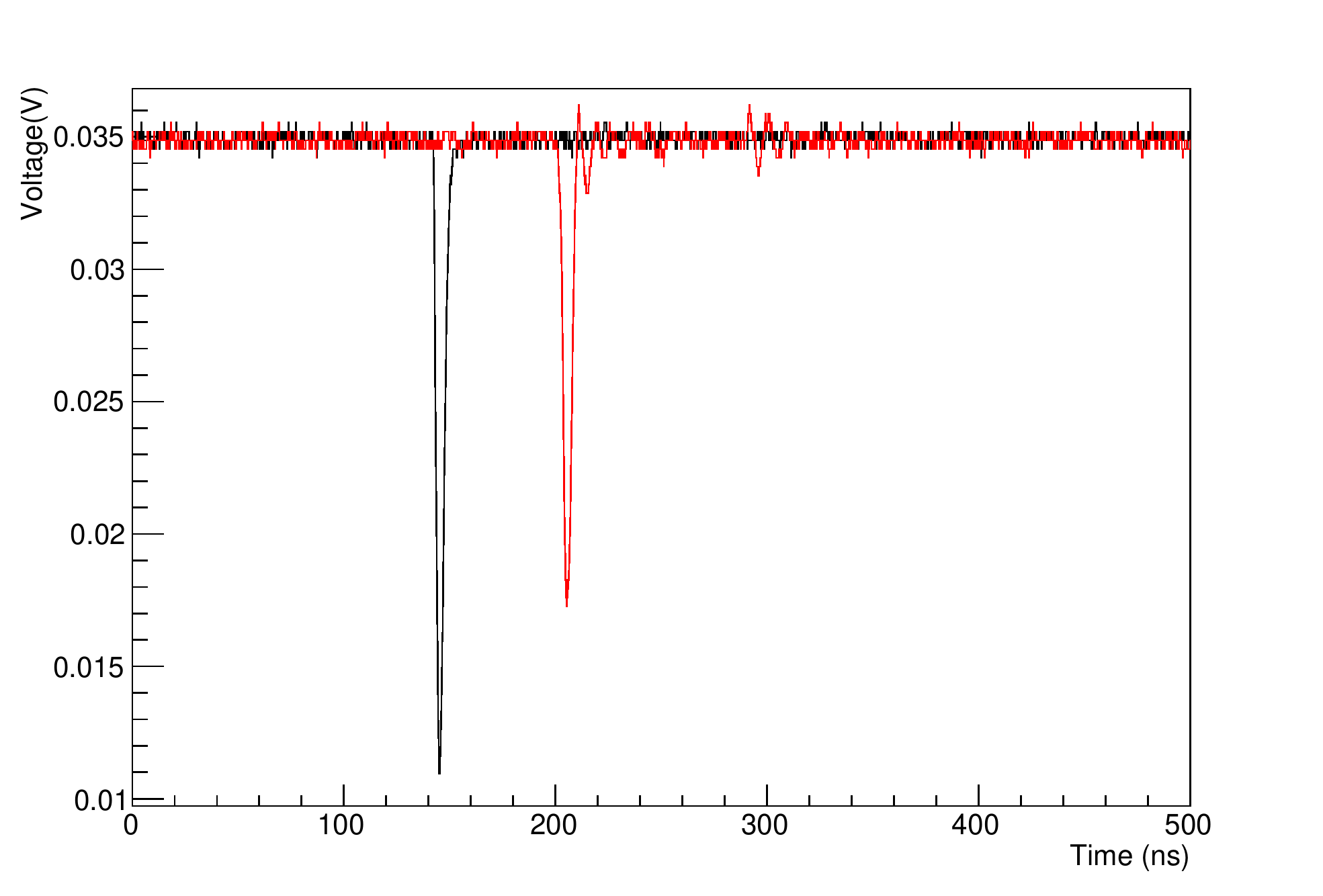}
\caption{The digitized waveforms for the R7600-200 trigger PMT (black) and the R5912-MOD PMT (red). The waveforms are 500.2ns long and are digitized in 50ps samples. The vertical and horizontal offsets were optimized for the analyses described in \ref{sec:charge} and \ref{sec:timing}.}
\label{fig:waveforms}
\end{figure}

Offline analysis code was used to find coincidences between the two PMTs, which are integrated and discriminated to determine charge and timing spectra, detailed in Sections \ref{sec:charge} and \ref{sec:timing} respectively. Figure \ref{fig:waveforms} shows the digitized waveforms for a coincidence between the trigger PMT and the R5912-MOD PMT. The coincidence rate is around 3\% due to the distance the R5912-MOD is kept from the source. One million waveforms are taken for each dataset in order to obtain around 30,000 coincidence events. This ensured that statistical uncertainties on most of the extracted SPE parameters are less than 1\%.

\subsection{Charge}\label{sec:charge}

The SPE charge distribution is an important characteristic of the PMT performance. The shape of the distribution largely describes the efficiency at which a SPE PMT pulse crosses a given discriminator threshold, which is referred to as the PMT channel efficiency in this paper. The parameter that most clearly indicates a high channel efficiency is the peak-to-valley (P/V) of the charge distribution. In these measurement the PMT was operated at a gain of $1 \times 10^{7}$ which is indicated by a peak in the charge histogram at 1.6pC.

\begin{figure}[ht]
\centering
\includegraphics[scale=0.8]{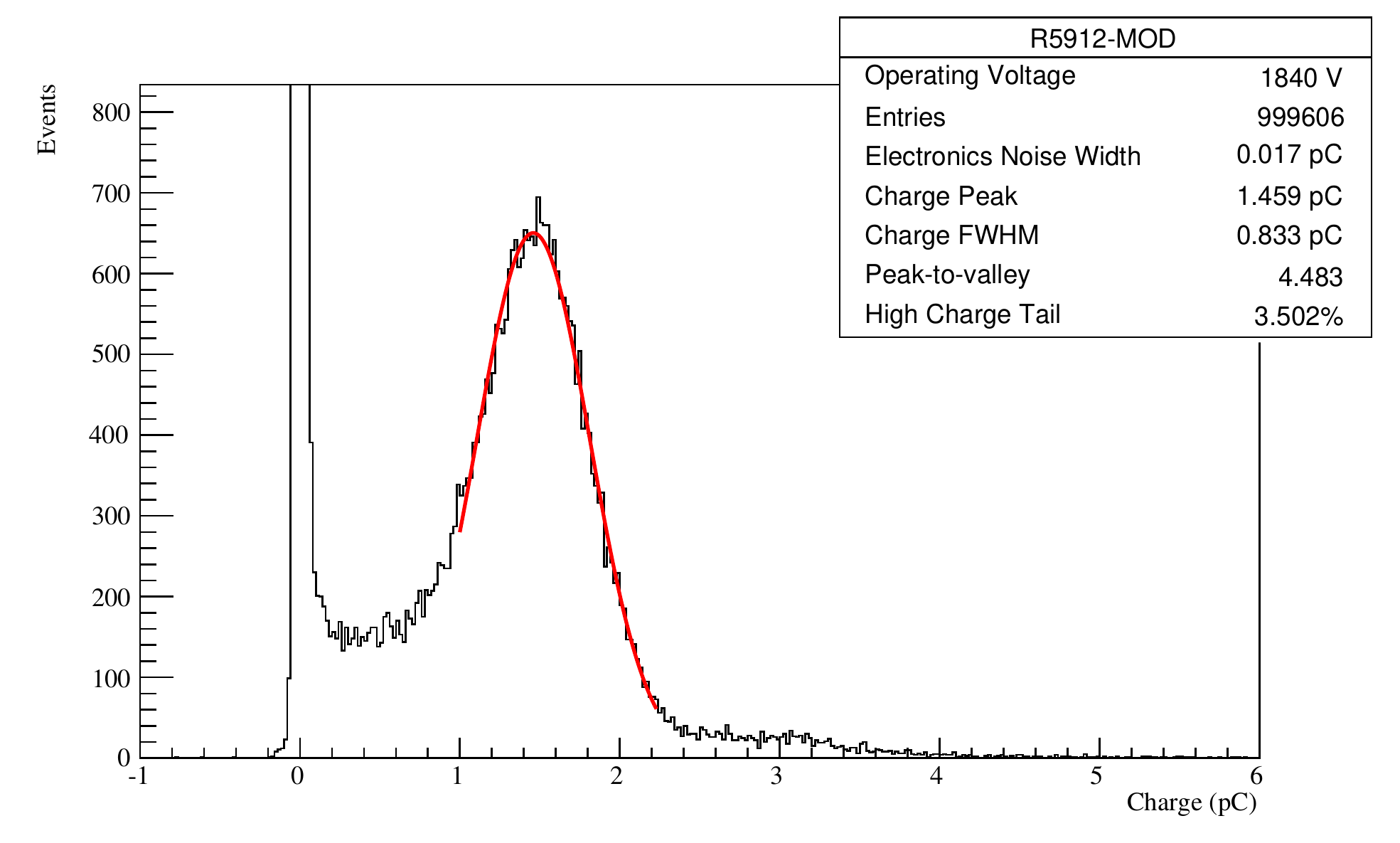}
\caption{The charge distribution of the R5912-MOD PMT. Shown in the statistics box is some important characteristics of the SPE charge response. The Guassian fit to the SPE peak is shown in red.}
\label{fig:charge}
\end{figure}

In order to produce the SPE charge distribution shown in Figure \ref{fig:charge}, the analysis code integrates each waveform using a 30ns window around the arrival time of the prompt light. The first 100ns of the waveform is used to calculate the baseline of the waveform. As seen in Figure \ref{fig:waveforms} the prompt light comes well after the baseline window has ended. However, on occasion a PMT pulse generated by dark current ends up in the baseline window, which drags the baseline down. If there is any pulse above the electronics noise in the baseline window the entire waveform is thrown out. This ends up rejecting much less than 1\% of the waveforms as can seen in Figure \ref{fig:charge} where the entries statistic shows that 999,606 of the one million total waveforms pass this cut. 

The prompt window is found by first looping over every waveform and calculating the average waveform of the entire dataset. Even at a relatively low coincidence rate, where most of the waveforms have no PMT pulse, the average waveform clearly indicates the location of the prompt light. Once the average waveform is found the prompt window is determined to be the window define by the $P - 12$ns to $P+ 18$ns where $P$ is the peak of the average waveform in ns. The SPE charge parameters are determined as follows. \\

{\bfseries Electronics Noise Width (ENW)}: The electronics noise is the large peak shown around 0pC in Figure \ref{fig:charge}. These noise entries come from integrating the PMT waveform for events with no coincidence pulse, which make up the majority of the waveforms. To extract the ENW a Gaussian is fit around the noise peak using the bins with values corresponding to half the peak height. It's important to keep the ENW to a minimum in order to ensure we can properly identify the valley of the charge distribution.

{\bfseries Charge Peak}: The bin above the ENW with the maximum content is identified. A Gaussian is fit between $\frac{2}{3}$ and $\frac{3}{2}$ of the charge value of this bin. The fit is shown in Figure \ref{fig:charge}. The mean of this fit is taken as the charge peak. The charge peak is an indication of the gain of the PMT, where a gain of 1.0$\times 10^{7}$ corresponds to a charge peak of 1.6pC. \\

{\bfseries Charge HWHM}: $2\sigma_{peak}\sqrt{2\log(2)}$, the FWHM as determined by the fit to the charge peak.

{\bfseries Peak-to-Valley (P/V)}: The height of the charge peak divided by the height of the minimum of the valley. The minimum of the valley is determined by a quadratic fit between the electronics noise and the charge peak. This parameter is a strong indicator of the PMT's channel efficiency. As will be discussed in Section \ref{sec:eff} this PMT has a very large P/V when comparing to other 8'' PMTs. 

{\bfseries High charge tail}: The number of events above $3\sigma_{peak}$ divided by the number of events above $3\sigma_{ENW}$. This indicates the amount of two PE contamination into the SPE sample. Once can clearly see in Figure \ref{fig:charge} the two PE peak around 3pC. The clarity of this peak is indicative of the extremely narrow charge distribution and excellent P/V. For most PMTs the 2 PE peak is washed out by a tail to the SPE distribution.

Overall the excellent SPE charge response of the R5912-MOD is characterized primarily by its narrow distribution and large P/V. Comparisons to other PMTs are made directly in Section \ref{sec:eff}.

\subsection{Transit Timing}\label{sec:timing}

The transit time is the amount of time it takes for a photoelectron created at the photocathode to travel through the PMT and be detected as an output pulse from the anode. This time varies from one photoelectron to the next and the spread in the transit time distribution is one of the most important characteristics of a PMT. The transit time spread is largely determined by the electronic optics of the PMT particularly between the photocathode and first dynode as well as between the first and second dynodes.

\begin{figure}[ht]
\centering
\includegraphics[scale=0.8]{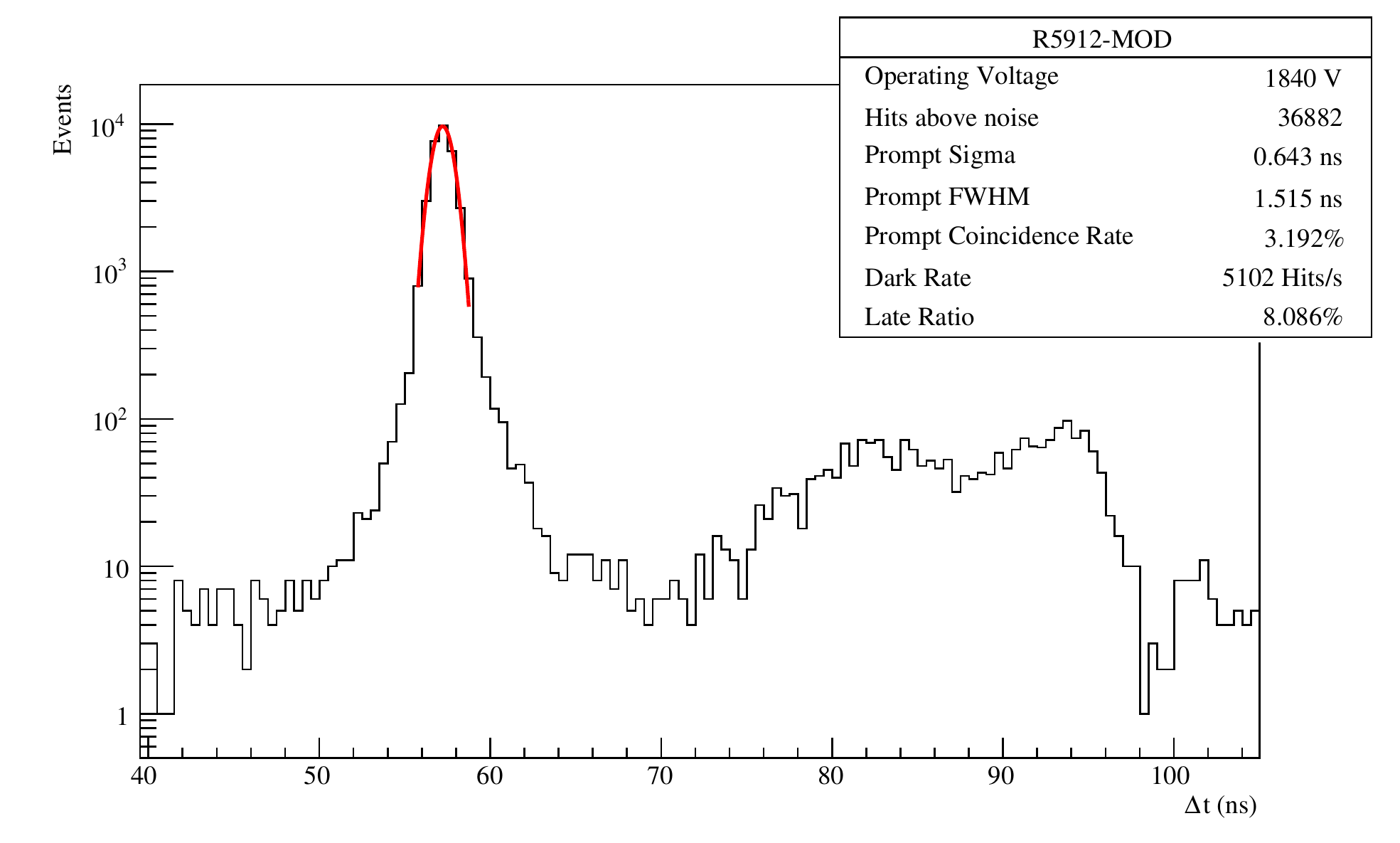} \\
\caption{The transit time profile of the R5912-MOD. Shown in the statistics box is some important characteristic of the SPE time response. The Gaussian fit to the prompt light peak is shown in red.}
\label{fig:timing}
\end{figure}

The analysis to extract the transit time distribution uses many of the same analysis techniques as described in Section \ref{sec:charge}, including extracting the baseline for each waveform. In order to extract the timing distribution, coincidence events are identified and the peaks of both the R5912-MOD PMT and the trigger PMT signals are found. Then a constant fraction discriminator is applied in analysis to each waveform and the samples corresponding to 20\% of the peak height are found. The time difference between those samples for the R5912-MOD and trigger PMT is found and that $\Delta t$ distribution as shown in Figure \ref{fig:timing}. As with the analysis described in Section \ref{sec:charge} the waveforms are sampled at 50ps.

To extract the charge distribution only the prompt light was considered. In order to extract the full transit time distribution the waveform is stepped through in 30ns windows, looking for PMT pulses. If the charge of the 30ns window is larger than 0.2pC the peak of the waveform is found and the $\Delta t$ is histogrammed. That means that a single waveform can contribute multiple times to the timing distribution. This happens most often for double pulsing and for waveforms that have both a dark pulse and a prompt pulse. This is described more later in relation to Figure \ref{fig:timing-types}. The 0.2pC threshold is determined by the width of the electronics noise so as not to accept electronics noise into the transit time histogram. The following timing distribution characteristics are extracted. \\

{\bfseries Hits above noise}: The number of PMT pulses in the transit time distribution. The total number of waveforms analyzed is 1 million. 

{\bfseries Prompt sigma}: The prompt light is the primary contribution to the transit time distribution. The spread in time of the prompt light is characterized by fitting around 10\% of the peak height on either side of the peak. The Gaussian fit is shown in Figure \ref{fig:timing} and the sigma of that fit, $\sigma_{SPE}$, is referred to as the transit time spread (TTS) in this paper. It should be noted that the spread in emission time of the Cherenkov light and the TTS of the trigger PMT both add negligible jitter to this measurement. The TTS extracted is influenced by several factors: the statistics in the peak, the systematics of the setup, the contamination of MPE hits, and uncertainties associated with the Gaussian fit to the peak. The number of waveforms in the datasets was intentionally taken in order to maintain around 1\% statistical uncertainty on the prompt peak. The systematics associated with the experimental setup and the multi PE contamination, which are tied together in the distance and angle from the source, were studied by taking data at various distance and angles from the source, up to a coincidence rate of roughly 5\%. The measured TTS varied by 2-3\%. Finally, the fit uncertainties were determined by running the fit over various reasonable ranges, rather than 10\% of the peak height on either side. This uncertainty turned out to be the largest at around 5\%. By fitting over smaller or larger ranges, the extracted TTS changes by around 20 - 30ps. Lastly it should be noted this parameter is sensitive to the PMT base used and with different bases to the one described in Section \ref{sec:base} we found TTSs up to 800ps. 

{\bfseries Prompt FWHM}: $2\sigma_{SPE}\sqrt{2\log(2)}$, the FWHM of the prompt fit.

{\bfseries Prompt coincidence rate}: The coincidence rate of the prompt light. This parameter was intentionally kept to less than 5\% by placing the R5912-MOD PMT 30cm from the Cherenkov source. That was done to minimize MPE contamination into the SPE sample.

{\bfseries Dark rate}: The rate of PMT pulses that fall outside the late pulsing and prompt pulsing regions. The various timing components are discussed more in Section \ref{sec:timing-compontents}. This rate ranged between 2 - 5 kHz for the three measured PMTs.

{\bfseries Late ratio}: The rate of PMT pulses that fall within a late window, between 10ns and 60ns after the prompt peak. These values were chosen empirically and are somewhat arbitrary, but are fairly typical for an 8-inch PMT. The late pulsing will be described in more detail in Section \ref{sec:timing-compontents}.

Overall the R5912-MOD shows an excellent SPE timing response, in particular a very narrow TTS.

\subsection{Timing Components}\label{sec:timing-compontents}

\begin{figure}[ht]
\centering
\includegraphics[scale=0.8]{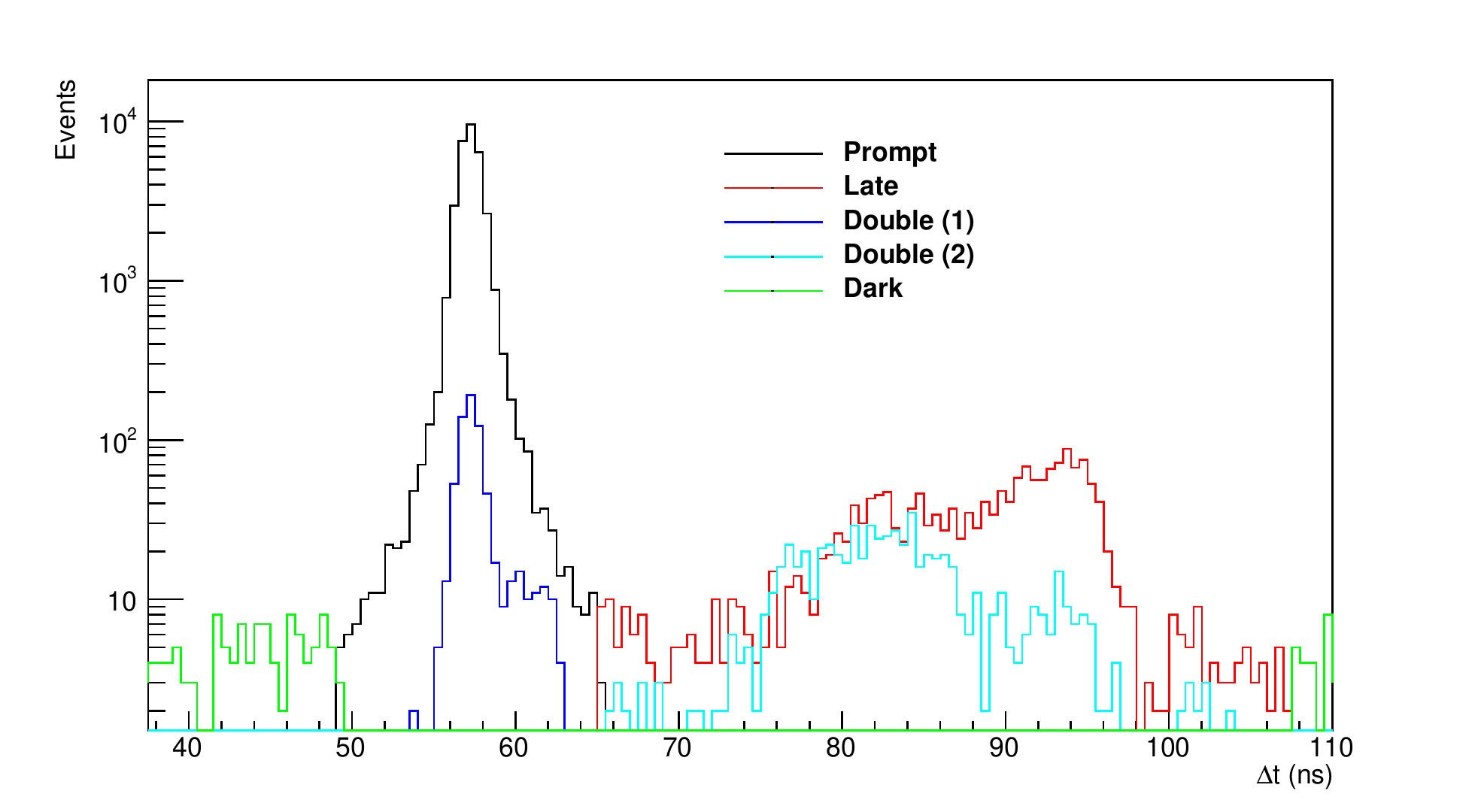}
\caption{The transit timing profile of the R5912-MOD broken down into the various components that make up the structure.}
\label{fig:timing-types}
\end{figure}

There are several different components that make up the PMT transit time distribution. In this paper we've considered prompt pulses, late pulses, double pulses, pre pulses, and dark pulses.  After pulses are considered separately in Section \ref{sec:afterpulsing}. In this section we discuss the physical model for each pulse type as well as the percent of the transit time distribution each pulse type makes up the R5912-MOD PMT. The timing breakdown of the various pulse types is shown in Figure \ref{fig:timing-types}. It should be noted the numbers presented are for measurement on the R5912-MOD PMT (model ZC2723) and the number varied by 1-2\% for the other two PMTs tested. 

{\bfseries Prompt pulses}: The prompt peak makes up the primary response of the PMT, the spread of which is influenced heavily by the electron optics in the PMT. The prompt peak makes up 91.7\% of the timing response. 

{\bfseries Late pulses}: The late light is the second largest component of the timing distribution, and is responsible for the peak at about 95ns in Figure \ref{fig:timing-types}. The late light is caused by an elastic scatter off of the first dynode in which the photoelectron that travels back toward the photocathode before returning to the first dynode and causing the emission of secondary electrons. The late pulsing for these PMTs makes up about 6.1\% of the PMTs response, which can be compared directly to the late ratio statistic in Figure \ref{fig:timing}, which does not correct for double pulsing. These late pulses are distinct from after pulses, which are caused by drifting ions in the PMT and are discussed in detail in Section \ref{sec:afterpulsing}.

{\bfseries Double pulses}: The double pulsing has a similar time structure to the late light; however, in addition to the late pulse there is also a prompt pulse in the waveform. The time structure of both the initial prompt pulse and the later pulse is shown in Figure \ref{fig:timing-types}. The double pulsing is caused by an inelastic scatter off of the first dynode. There is enough energy transfer for secondary emission to take place; however, the photoelectron also recoils back toward the photocathode. Because the initial photoelectron does not maintain its full energy, it recoils over a shorter distance and thus it is expected that second of the double pulses to come slightly earlier than the late pulses. Figure \ref{fig:timing-types} shows that is indeed the case; the second pulse in the double pulsing arrives early on average than the late pulsing. Double pulsing makes up about 2.2\% of the total timing structure.

{\bfseries Pre pulses}: Pre pulsing is caused when a photon is transmitted, rather than absorbed or reflected, by the PMT glass and photocathode. The photon travels through the PMT vacuum and can strike the first dynode, causing the creation of a photoelectron at the first dynode rather than at the photocathode. One would expect to see this pre pulsing peak in the transit time distribution about 10ns before the prompt peak; however we cannot resolve any pre pulsing above the dark rate. With our statistics we can claim pre pulsing makes up less than $0.1$\% of the total transit time distribution.

{\bfseries Dark rate}: Dark pulses are caused primarily by thermionic emission of an electron at the photocathode and are not caused by incident light. There are several other ways to get dark current at the photocathode, including Cherenkov light produce by muons passing through the glass. Dark pulses are not a part of the transit time distribution, but provide a flat background that is accounted for when calculating the various percent contributions. For the three PMTs tested the dark rate was between 2 - 5 kHz. We let the PMTs cool down for several hours in a dark box before extracting the dark rate. 

\subsection{PMT Waveform}

The shape of the SPE waveform is an important part of the PMT model. A lognormal distribution is often used to model the shape of PMT waveforms. However, given the excellent resolution of our DAQ the sum of three lognormals, described by Equation \ref{eq:lognormal}, is fit to the PMT pulses.

\begin{equation}\label{eq:lognormal}
f = \sum_{i=0}^{2}\frac{N_{i}}{t\sigma_{i}\sqrt(2\pi)} e^{-\frac{1}{2\sigma_{i}^{2}}\log({\frac{t}{\tau_{i}}})^{2}}
\end{equation}

\begin{figure}[ht]
\centering
\includegraphics[scale=0.4]{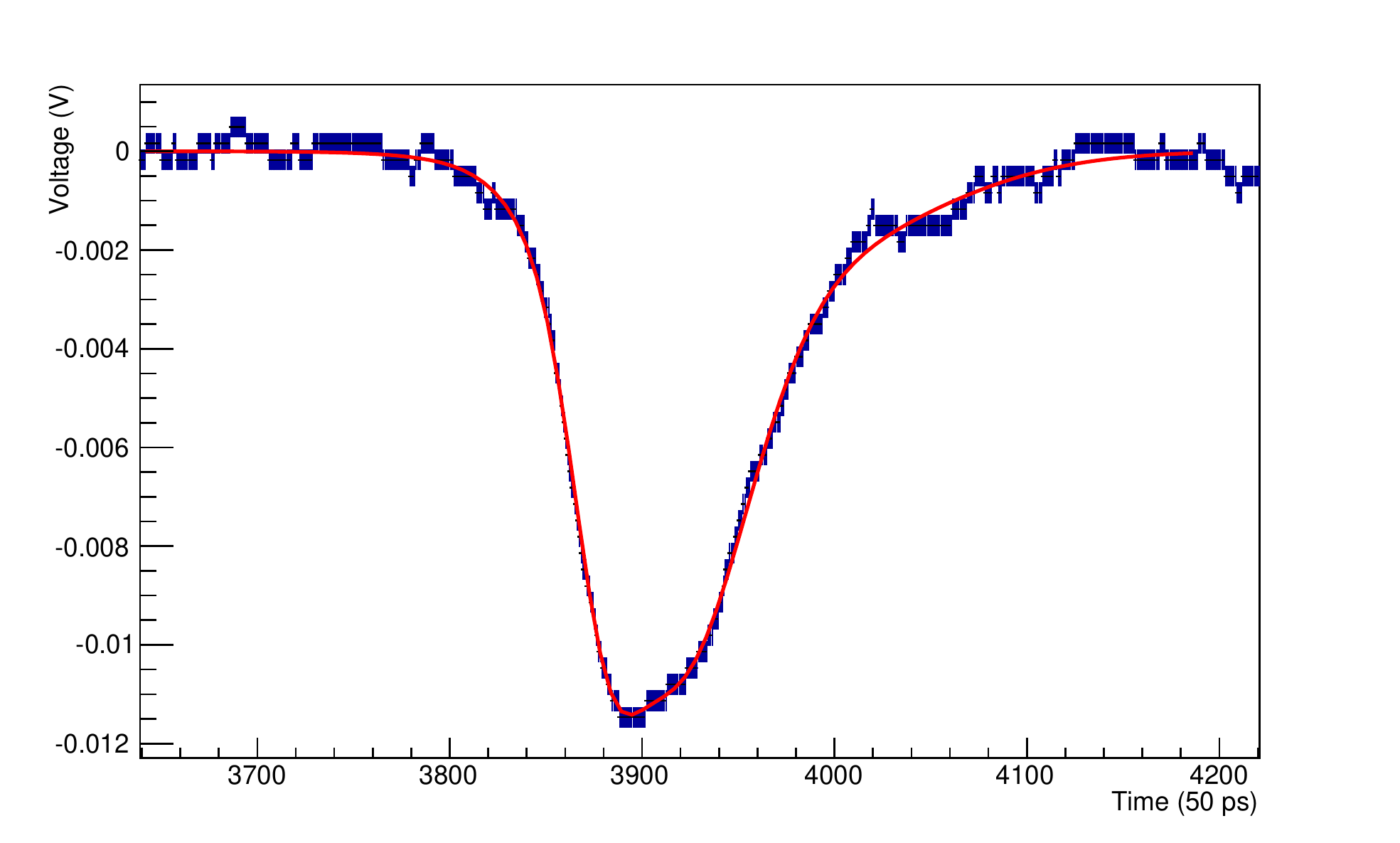}
\includegraphics[scale=0.4]{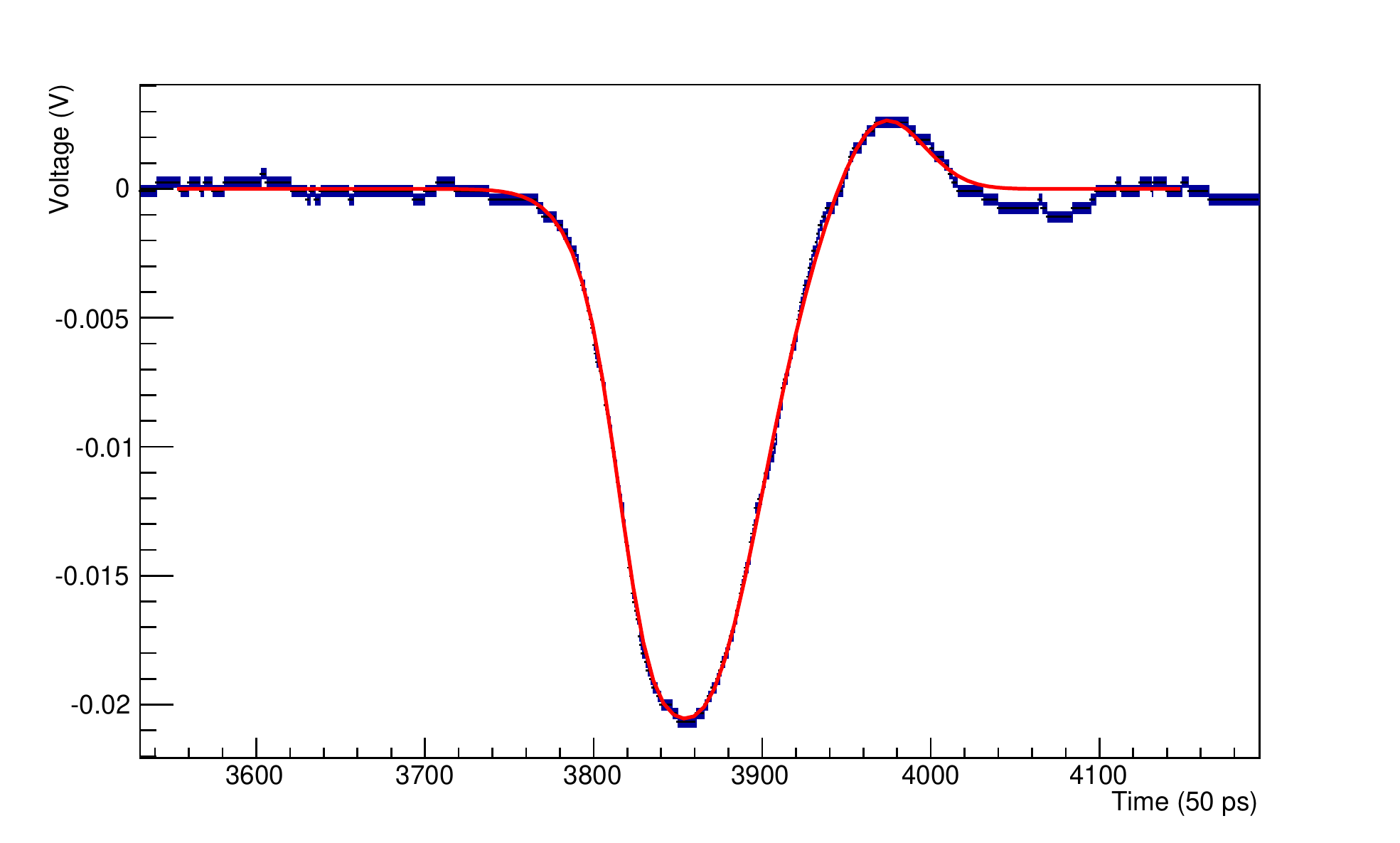}
\caption{Two examples of R5912-MOD PMT waveforms with a triple lognormal fit shown in red.}
\label{fig:lognormal}
\end{figure}

This improved model allows one to fit waveforms that have additional components to the PMT shape. Two such waveforms with the associated fits are shown in Figure \ref{fig:lognormal}. The triple lognormal distribution allows one to extremely accurately characterize the PMT shape, including the rise time, the fall time, the overshoot on the falling edge, and any additional structure in the waveform. It should be noted that the shape of the PMT waveform is in part determined by the base design discussed in Section \ref{sec:overview}.

\subsection{Performance Across Gains}

PMTs are operated across a wide range of gains. For most of the measurements in this paper the PMT is operated at a gain close to $1 \times 10^{7}$. However, by increasing the high voltage supplied one could operate the PMT at an even higher gain and improve the efficiency at which one detects PMT pulses that would otherwise not have crossed threshold. Shown in Table \ref{tab:gain} is the prompt coincidence rate, the peak, and the TTS of the ZC2723 R5912-MOD PMT across supply voltages ranging between 1700 and 2100V. As can be determined by comparing the values of the charge peaks, the corresponding gain of the PMT ranged from $0.45 \times 10^{7}$ to $1.7 \times 10^{7}$. One could imagine going to even higher PMT gains, but the design of the base limited the supply voltage to a little over 2000V. The ratio of the prompt coincidence rates is a good indicator of the relative efficiency gain from increasing the PMT gain. The gain in efficiency by increasing the high voltage from 1700 to 2100 is about 12\%. For our setup and DAQ, the change in efficiency is primarily due to PMT pulses that fall into the electronics noise, as described in Section \ref{sec:charge}. As the gain increases fewer PMT pulses are lost in the electronics noise. It is also possible that the collection efficiency, which indicates the efficiency at which a photoelectron created at the photocathode successfully reaches the first dynode, could be improving with the higher supply voltage. We did not attempt to disentangle these effects.

The TTS has very little dependence on the gain of the PMT. The small amount of jitter seen across the gains is consistent with the size of the fit uncertainties, as discussed in Section \ref{sec:timing}.

\begin{table}[ht]
\centering
\begin{tabular}{|c|c|c|c|}
\hline
High Voltage (V) & Coinc. Rate (\%) & Peak (pC) & TTS (ns) \\ \hline \hline
1700 & 3.72 & 0.73 & 0.67 \\ \hline 
1760 & 3.82 & 0.91 & 0.64 \\ \hline 
1800 & 3.91 & 1.06 & 0.64 \\ \hline
1880 & 4.10 & 1.36 & 0.65 \\ \hline 
1920 & 3.99 & 1.55 & 0.65 \\ \hline
1960 & 4.12 & 1.78 & 0.62 \\ \hline
2000 & 4.10 & 2.01 & 0.62 \\ \hline
2100 & 4.15 & 2.69 & 0.65 \\ \hline
\end{tabular} 
\caption{The SPE performance for the ZC2723 R5912-MOD PMT across gains spanning roughly $0.45 \times 10^{7}$ to $1.7 \times 10^{7}$. The coincidence rate changes by about 12\% across the gain change and the TTS is constant within uncertainties.}
\label{tab:gain}
\end{table} 

\section{Relative Efficiency and PMT Comparisons}\label{sec:eff}

In this section the efficiency of the R5912-MOD PMT is compared directly to the R1408 and R5912-100 PMTs, both of which are 8'' PMTs developed by Hamamatsu. The R1408 and R5912-100 have been used in large neutrino and dark matter detectors such as SNO \cite{SNO}, MiniBooNE \cite{MiniBooNE}, and DEAP \cite{DEAP-R5912}. The expected detection efficiency, as determined by the combinations of the quantum efficiency, collection efficiency, and channel efficiency, of the R1408 is close to 15\% and the R5912-100 is around 30\%. Three R5912-MOD PMTs and one of each of the R1408 and R5912-100 PMTs were characterized using the setup described in Section \ref{sec:setup}. All tests were performed under the same conditions, most importantly, the distance to the Cherenkov source was carefully controlled for. Table \ref{tab:pmt-rel} shows the comparison of several important SPE parameters of the R5912-MODs as compared to the measured R1408 and R5912-100. 

Given the primarily SPE response of the PMTs, the coincidence rates indicates the relative efficiency of the PMTs. In the second column of Table \ref{tab:pmt-rel} the coincidence rates have been normalized to the R5912-100 and show that R1408 at less than 50\% of the detection efficiency, as expected. The R5912-MODs are all around 65\% of the detection efficiency, suggesting around a 20\% overall detection efficiency. This lower efficiency is expected based on conversations with Hamamatsu. It should be noted no correction for differences in the QE curves was made, nor any correction for the different shapes of the charge distributions, which change the channel efficiency for our DAQ. However, given the relatively similar QE shapes, the broad wavelength spectrum of the Cherenkov light, and the very narrow ENW, these effects should be very small.

Perhaps most importantly, the TTS of the R5912-MOD outperforms the other PMTs. ZC2723 has the narrowest spread in timing of the three R5912-MODs measured which is more than twice as narrow as the R1408. The P/V is also superior to the compared PMTs, more than four times better than the R1408. The TTS and P/V of the R5912-100 also outperforms other modern large-area PMTs. As described in \cite{r11780} the Hamamatsu 12'' R11780 HQE PMT has a TTS of around 1.0ns and P/V a little over 2.0. The Double-Chooz experiment made an in-situ measurement of the TTS of the R7081 Hamamatsu 10'' PMTs of 0.9ns \cite{r7081}, similar to what was found to the R5912-100 in this paper. The 11'' D784UKFLB PMTs designed by ET Enterprises and planned to be used in the Annie detector \cite{ANNIE} were measured to have transit time spreads around 2.0ns. In terms of SPE timing, the R5912-MOD outperforms all of these PMTs.

\begin{table}[ht]
\centering
\begin{tabular}{|c|c|c|c|c|c|c|}
\hline
PMT & Coinc. Rate (\%) & Rel. Eff. & TTS (ns) & Late Ratio (\%) & P/V & HV \\ \hline \hline
R5912-HQE & 5.32 & 1.0 & 0.87 & 7.58 & 2.96 &  1740V \\ \hline
R1408 PBUT & 2.47 & 0.46 & 1.51 & 7.18 & 1.08 & 2000V \\ \hline
R5912-MOD ZC2722 & 3.47 & 0.65 & 0.72 & 8.34 & 4.28 & 1840V \\ \hline
R5912-MOD ZC2723 & 3.44 & 0.65 & 0.63 & 8.09 & 4.46 & 1940V \\ \hline
R5912-MOD ZC2728 & 3.39 & 0.64 & 0.69 & 8.94 & 3.96 & 1740V \\ \hline
\end{tabular}
\caption{The table of coincidence rates, TTS, late ratios, and peak to valleys of the three R5912-MODs tested as well as an R1408 and R5912-100 for comparison.}
\label{tab:pmt-rel}
\end{table}

\section{1D Scan}\label{sec:1d}

A simple 1D scan of the PMT was performed in order to roughly characterize the response across the front face of the PMT. The measurements described in Section \ref{sec:SPE} probe the entire surface area of the photocathode by moving the PMT far enough away from the source. However, it is well known that the PMT performance diminishes at the edges of the photocathode, primarily due to the electron optics between the photocathode and the first dynode. In order to perform the scan the PMT was mounted in a fixed orientation. The Cherenkov source described in Section \ref{sec:setup} was masked off using black electrical tape and a black plastic mask with an approximately 1cm diameter hole drilled in the center. The size of the hole was carefully determined in order to achieve a roughly 3\% coincidence rate, similar to what was achieved by keeping the PMT 30cm from the source. The masked off source was then pressed against the front face of the PMT so as to probe only a small portion of the photocathode area. The incident angle was controlled for by ensuring the source was pushed flush to the PMT. The edge of the photocathode was taken from Figure \ref{fig:pmt} where it is shown to extend to $\phi = 100$mm from the center.

With this setup, the source was moved around the PMT in 10 degree steps from one edge of the photocathode to the other. The systematic uncertainties were determined by repeating the measurement several times at each angle, and were found to be the largest source of error. These are probably driven by relatively large uncertainties in the precise angle between the source and the center of the PMT. By repeating the full scan of the measurements several times systematic uncertainties of 8\% were found by looking at the variation of the measured parameters. The same statistical and fit uncertainties described in Section \ref{sec:SPE} hold true for this measurement. 

\begin{figure}[ht]
\includegraphics[scale=0.4]{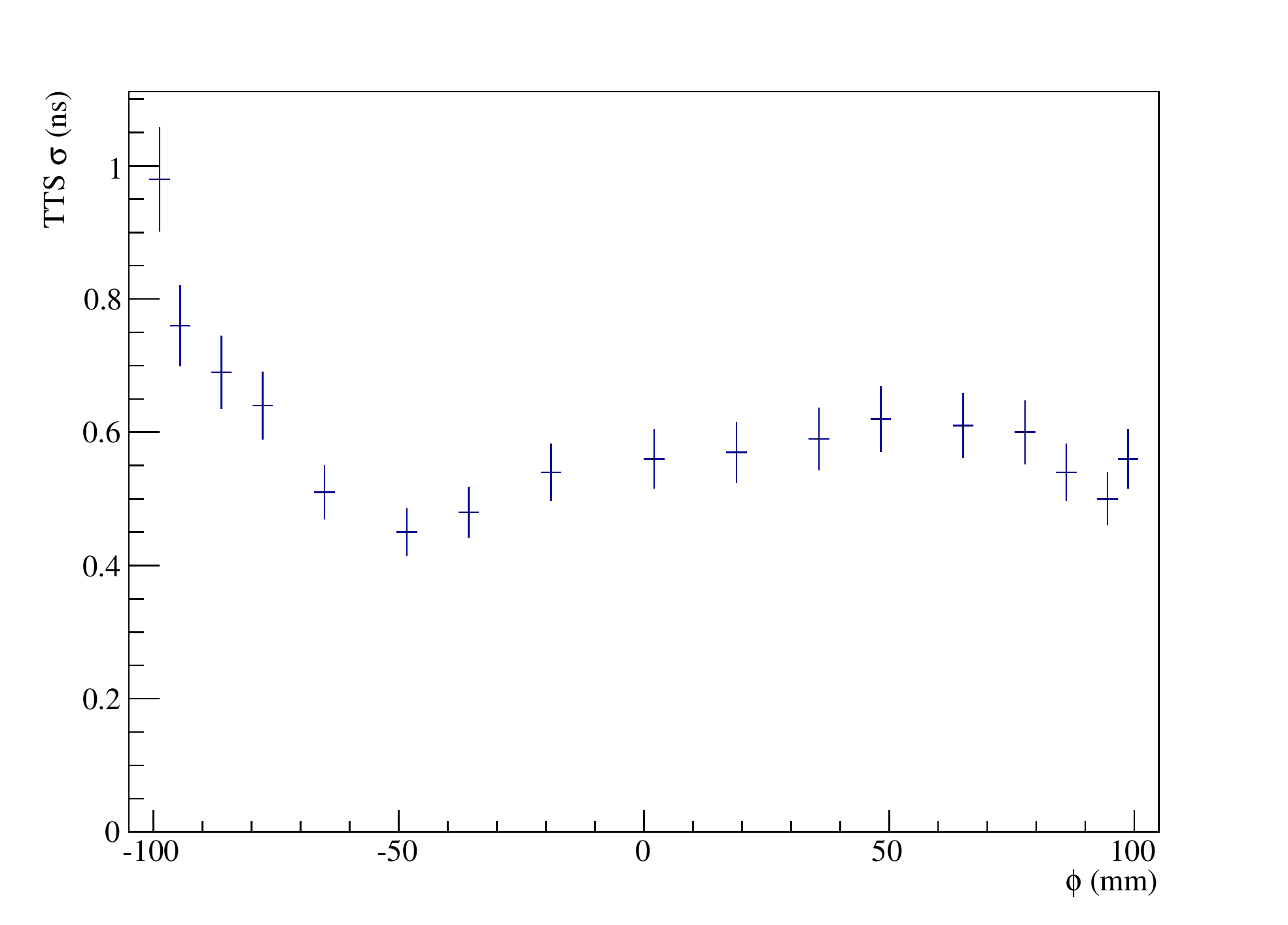}
\includegraphics[scale=0.4]{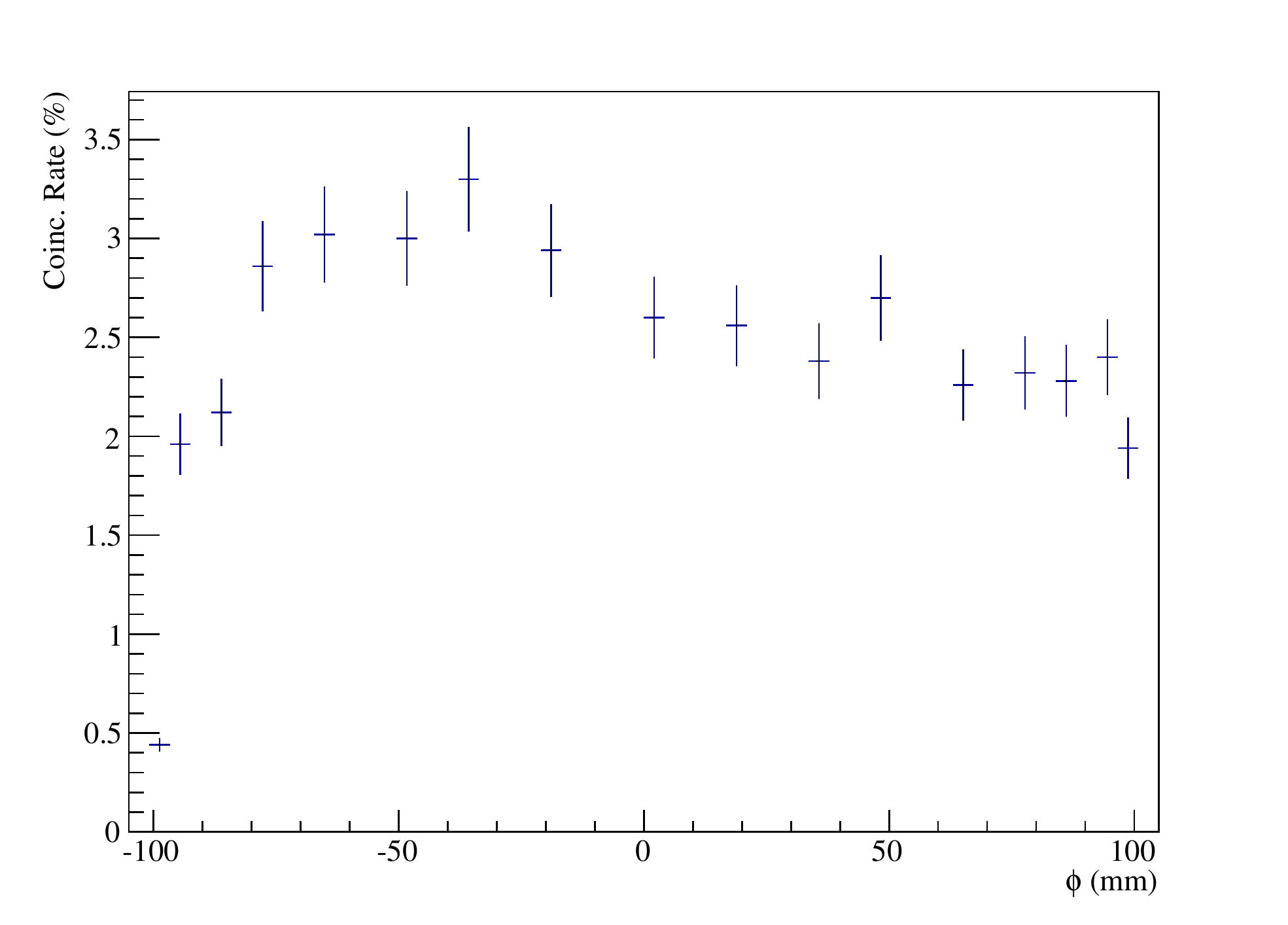}
\caption{The TTS (left) and the coincidence rate (right) of the R5912-MOD PMT as a function of $\phi$, the perpendicular distance to the center of the PMT.}
\label{fig:1D}
\end{figure}

The TTS and the coincidence rate of the PMT hits is shown as a function of $\phi$ in Figure \ref{fig:1D}. These parameters were extracted precisely the same way as described in Section \ref{sec:timing}. This simple scan makes clear the increase in TTS and decrease in efficiency at the edge of the photocathode; however, for this particular scan one edge was significantly worse than the other. This lack of symmetry is not untypical of PMTs and can be seen more explicitly for a 12'' PMT in \cite{r11780}. The peak of the charge distribution, was flat in $\phi$ across the scan, suggesting the gain of the PMT is impacted little by the location of the photocathode the photon strikes. Additionally, it should be noted that at several places along the scan the $\sigma$ of the TTS dropped below 0.5ns, an impressive feature of these PMTS. An average of the TTS over all the measurements shown in Figure \ref{fig:1D} gives an average TTS of 0.63 $\pm$ 0.05, which is consistent with the TTS of 0.64ns found in Section \ref{sec:timing}, where the setup illuminated the entire PMT.

As discussed already the prompt coincidence rate is a good indicator of overall efficiency, and the coincidence rate plot in Figure \ref{fig:1D} shows a fairly flat efficiency almost all the way out to the edges. On one outer edge the efficiency did fall by about a factor of five, but the fact that the efficiency did not drop drastically until the very edge of the photocathode is promising.

Overall the 1D scan, while not a precise determination of the 2D performance of the PMT, gives an idea how some of the important SPE parameters change as a function of photocathode position. This measurement has much room for improvement, and future work would work to improve the setup to precisely measure the 2D response of the PMT, as was done in \cite{r11780}. 

\section{After pulsing}\label{sec:afterpulsing}

After pulsing is an additional component to the PMT timing that can arrive server $\mu$s after the prompt light. There exists a small amount of residual gas, most commonly Helium, in the PMT vacuum that can be ionized by the photoelectrons as they travel from the photocathode. These positive ions then drift back to the photocathode, which they strike and produce electrons. These secondary electrons act just like photoelectrons and create PMT pulses up to several microseconds after the initial photoelectron.

After pulsing is an important part of the PMT model, especially because many experiments care about PMT hits on the time scale of $\mu$s or longer. Direct dark matter experiments often use the long triplet tail of the scintillation light in order to do pulse shape discrimination. PMT afterpulsing can look like late hits in the scintillation tail, so it's critical to model properly. For many liquid scintillator experiments, $\beta$-$\alpha$ separation relies on late PMT hits 100s of nanosecond after the prompt light, which can be effected by the faster after pulses.

\subsection{Experimental Setup}

The setup used to probe after pulsing is fairly simple. A 390nm LED is collimated and passed through a 10nm wide optical filter, directed at the center of the R5912-MOD PMT. The beam spot of the LED is directed at the center of the PMT and is tuned to be about 2cm in diameter. The signal to the LED is provided by a Agilent 33503A waveform generator, which pulses the LED with roughly 30ns wide square pulse. The intensity of the LED is tuned using the pulse amplitude, and measurements have been made at several different intensities. The frequency of the pulses is set to 1kHz so that there is no pile-up.

The same DAQ system as described in Section \ref{sec:setup} is used, with only one major difference. Rather than triggering on the R7600-200 PMT, the signal from rising edge of the signal provided by the waveform generator is used as a trigger. This ensures that the prompt light at the PMT comes after the trigger at a fixed amount of time. The scope settings are also adjusted so that the waveforms extracted are much longer in length, 50$\mu$s, and the sampling time used is 0.1ns which provides far better resolution than necessary. There is an intrinsic jitter on the prompt signal at the PMT associated with the width of the pulse driving the LED, which is dealt with in analysis. Because the after pulsing distribution is very broad in time this 30ns is a negligible jitter.

\subsection{Results}

\begin{figure}[ht]
\centering
\includegraphics[scale=0.8]{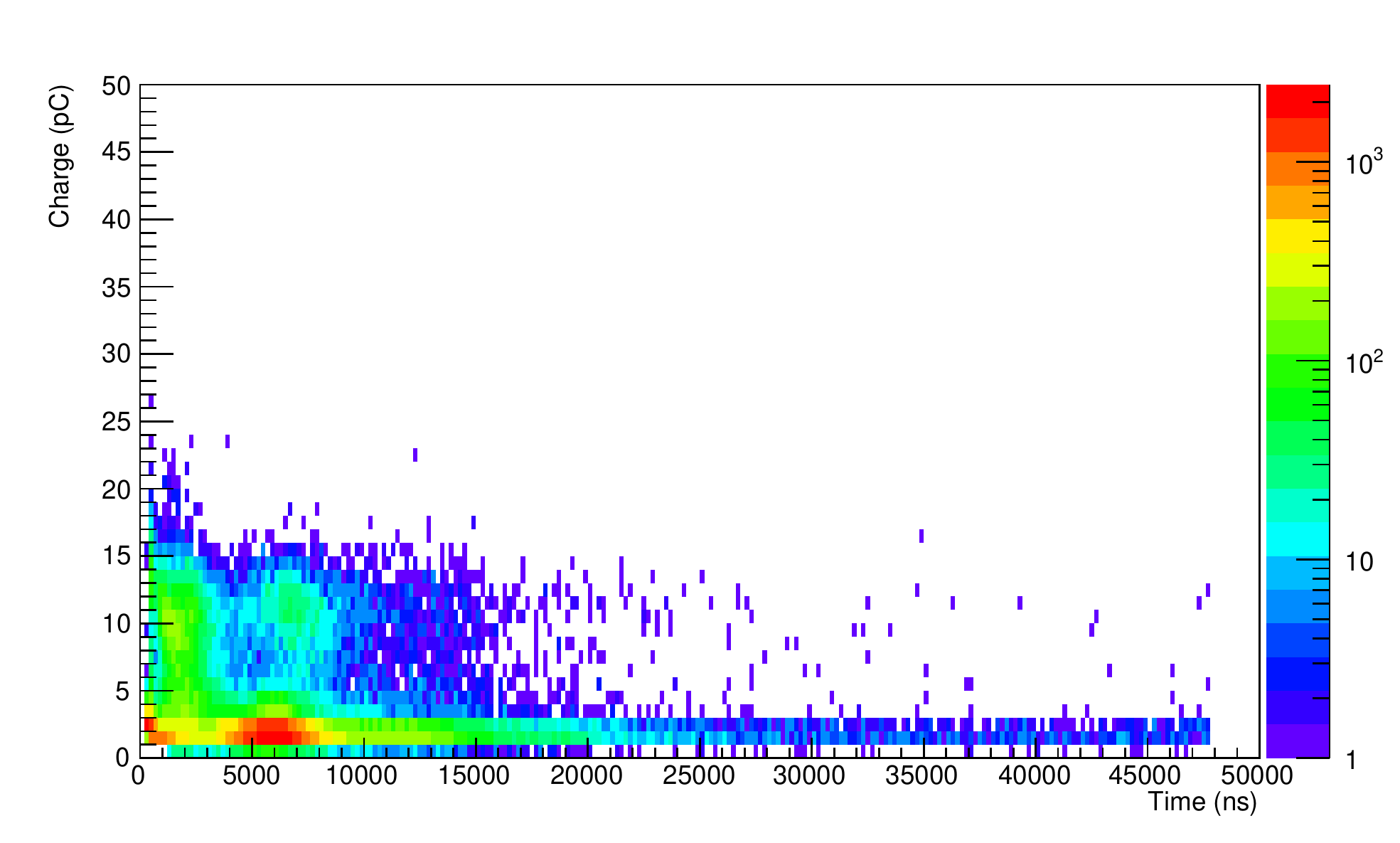}
\caption{A 2D histogram showing the time between the prompt light and the after pulse plotted against the charge of the after pulse.}
\label{fig:afterpulsing}
\end{figure}

In order to extract the time of the after pulse hits, the waveforms of the R5912-MOD are analyzed using similar techniques as described in Section \ref{sec:SPE}. The first 1$\mu$s of the window is used to calculate the baseline of the waveform and the scope setting are set so that this window does not contain any light from the LED. The rising edge of the prompt light is found and taken as t$_{0}$. This prompt light is then integrated over a 200ns window in order to calculate the charge in the prompt window. This 200ns window was found to contain all of the `prompt' light which includes late pulses and a jitter from the function generator. After the 200ns window, the rest of the 50$\mu$s waveforms is stepped through in 40ns windows, where any pulse crossing a threshold of 4mV is counted as an after pulse. Additionally a condition that the pulse must be broader than 5ns is applied in order to reject spikes in electronic noise. The 4mV chosen is just above the electronic noise for this measurement. The peak of the pulse is found and the time is taken as the time of the 20\% peak-height crossing minus t$_{0}$. Additionally, the charge of the after pulse is found by integrating around the pulse in a 30ns window. Figure \ref{fig:afterpulsing} shows a 2D histogram of the time vs. charge of the after pulses identified. There are two broad peaks around 1$\mu$s and 6$\mu$s with some additional structure early in time (around 400-600ns) and late in time (around 10$\mu$s). The PMT for this measurement was run with a gain such that the peak of the charge distribution was around 2pC. Figure \ref{fig:afterpulsing} shows that the majority of the after pulsing are SPE, with a fairly large tail out to about 10 PE. The flat distribution after 20$\mu$s is primarily due to dark pulses.

\begin{figure}[ht]
\centering
\includegraphics[scale=0.8]{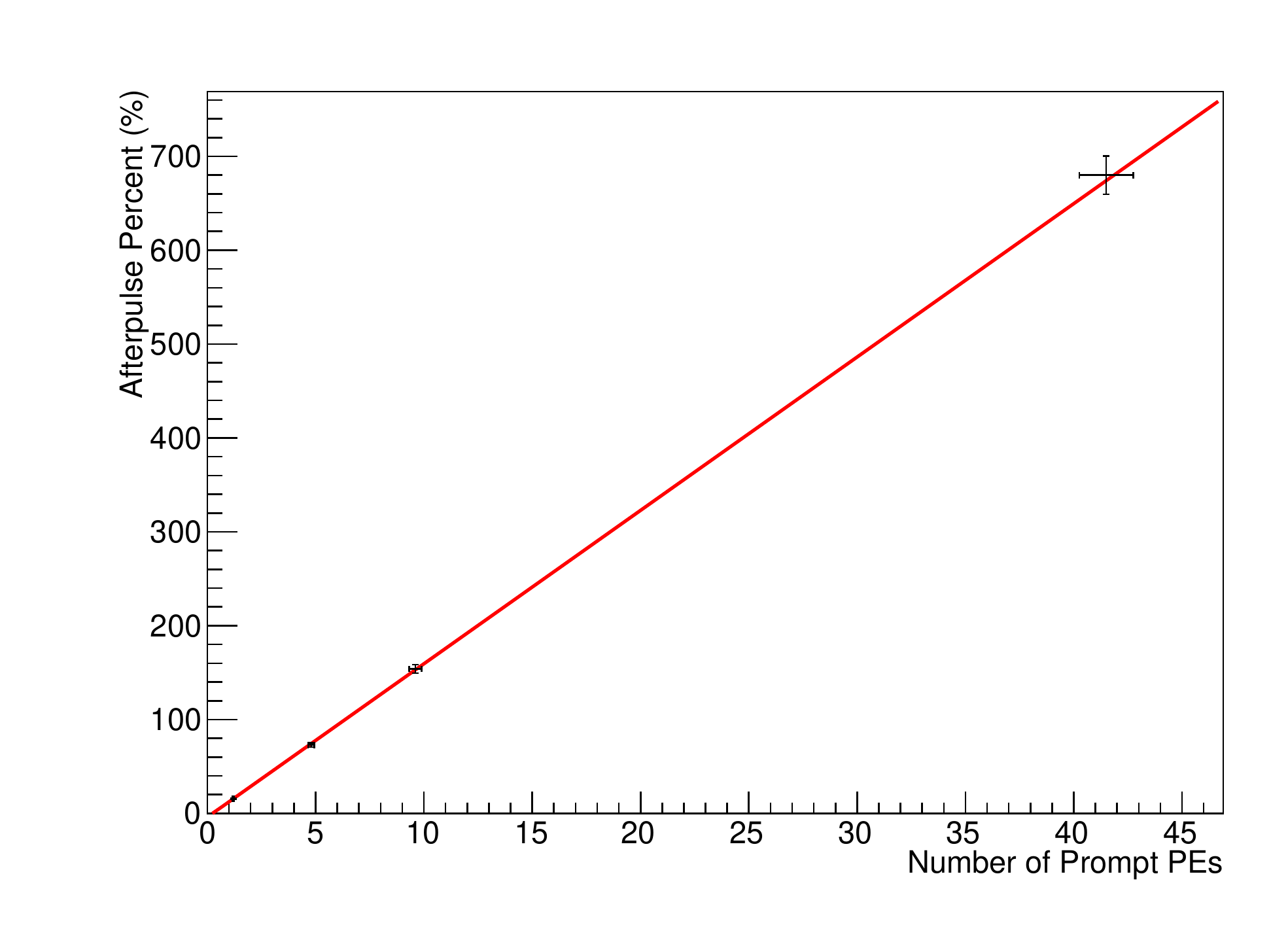}
\caption{The percent change of detecting an after pulse as a function of the number of prompt photoelectrons. A percentage larger than 100\% indicates multiple after pulses were detected on average. The linear fit shown fit to the data predicts a 16\% after pulsing probability per prompt photoelectron.}
\label{fig:appct}
\end{figure}

As mentioned, the after pulsing data was taken for multiple LED intensities in order to probe the probability of an after pulse, given an initial number of prompt PEs. Figure \ref{fig:appct} shows the percent change of getting an after pulse as a function of the number of prompt photoelectrons. The after pulsing probability calculation is fairly simple. The number of events identified as an after pulse, corrected by the dark rate of the PMT, divided by the number of pulses sent by the function generator. The intensity of the LED is high enough so that light is detected for every pulse from the function generator. A percentage larger than 100\% means that on average multiple after pulses were found in the waveform.

\begin{figure}[ht]
\centering
\includegraphics[scale=0.8]{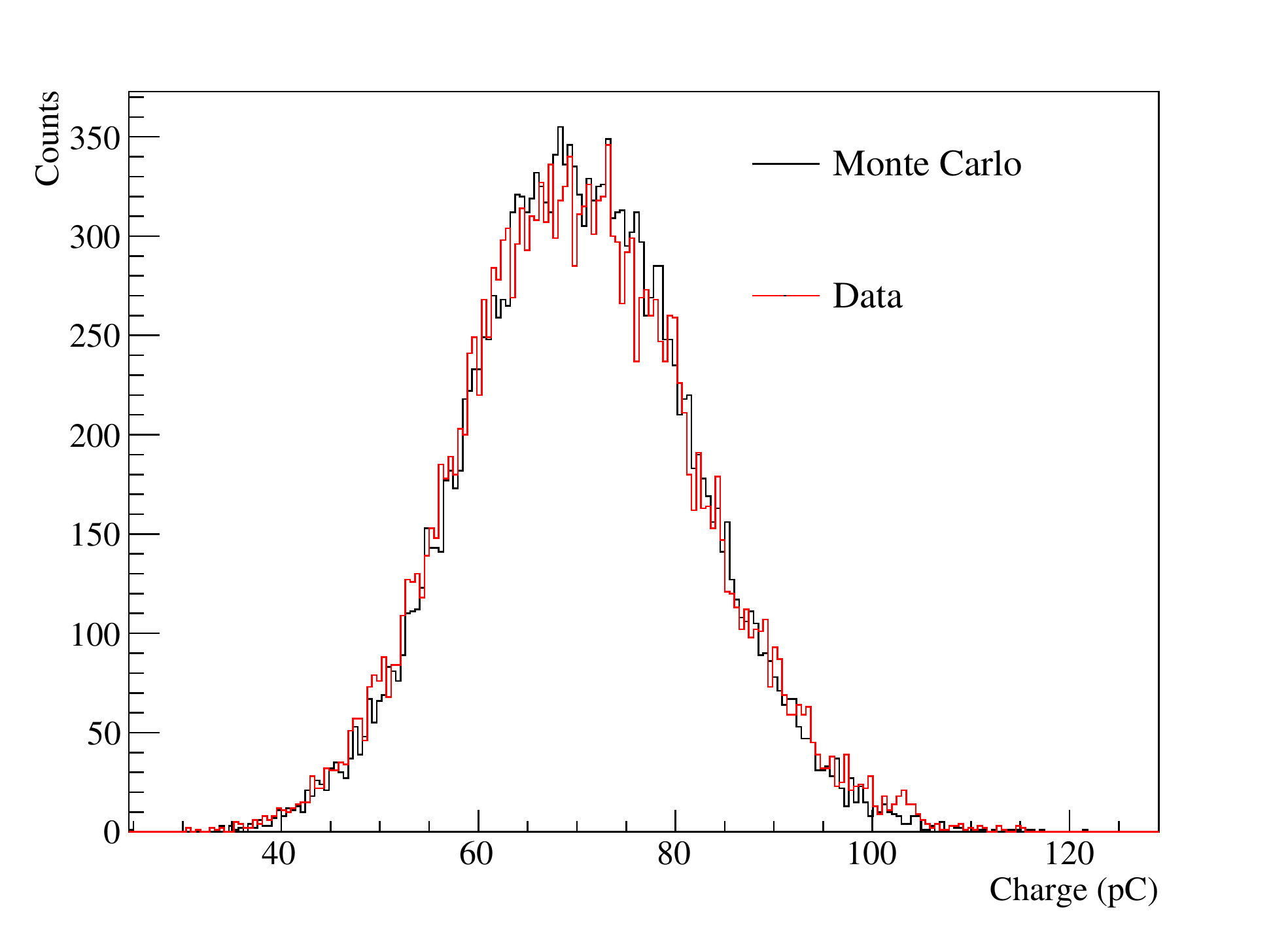}
\caption{The charge of the prompt light compared against the toy MC, used to extract the number of prompt PEs for Figure \ref{fig:appct}.}
\label{fig:npe}
\end{figure}

The calculation to determine the number of after pulses in the prompt-window is slightly more involved. First the SPE charge distribution, just like the one shown in Figure \ref{fig:charge}, is extracted for the PMT at the appropriate gain. The SPE distribution is then convolved with a Gaussian with a mean of $N_{PE}$ and a width of $\sqrt{N_{PE}}$, where $N_{PE}$ is the number of prompt PEs. That convolved distribution is then compared against the extracted charge distribution of the prompt light. The value of $N_{PE}$ is then tuned so that this toy MC agrees most closely with data, which is characterized by a $\chi^{2}$. Figure \ref{fig:npe} shows an example of the toy MC compared directly to the data, for an extracted $N_{PE}$ of 41.5. This particular example was the highest intensity setting we used, and is also shown on Figure \ref{fig:appct}. 

Figure \ref{fig:appct} shows four after pulsing measurements made at different intensities. The success of the simple fit shown indicates that the percent of after pulsing is linear across a broad range of intensities. The slope of the line measures the percent change of after pulsing per prompt photoelectron, which is found to be 16\%. This is a very high after pulsing probability. An identical measurement for the R1408 showed an after pulsing probability of around 1\% per prompt photoelectron. The probability for after pulsing for the R5912-100 was characterized in-situ by the DEAP experiment \cite{DEAP-R5912} and was found to be slightly less than 1\%. Given that these are prototype PMTs,  Hamamatsu might not have taken extra care taken to minimize residual gas in the PMT vacuum; however, the authors can only speculate at this point regarding the particularly high amount of after pulsing measured. 

\section{Conclusion}
Overall the R5912-MOD are excellent candidates for future optical detectors particularly because of their narrow spread in the prompt timing and large peak to valley ratio of the charge distribution. Ideally, one would combine the excellent SPE response with the high quantum efficiency of the R5912-100 PMTs. The SPE parameters for the R5912-MOD were directly compared to an R5912-100 and R1408 PMTs in the same experimental setup and were either better than or consistent with these PMTs. The 1D scan showed that the R5912-MOD TTS and efficiency gets worse at the edge of the photocathode, but it also showed that near the center of the PMT the spread in the transit time dropped to around 0.5ns. Finally, the R5912-MOD PMT after pulsing measurement showed a particular high after pulsing probability.

\section{Acknowledgements}
This work was supported by the Department of Energy, the Office of Nuclear Physics, and the University of Pennsylvania. The author would like to thank Josh Klein for providing important feedback and PMT expertise and Alexandra Ulin for her hard-work on the 1D scan measurements.

\end{document}